\documentclass{article}

\PassOptionsToPackage{numbers,sort&compress}{natbib}
\usepackage[final]{tackling_climate_workshop_style}
\usepackage[utf8]{inputenc}
\usepackage[T1]{fontenc}
\usepackage{hyperref}
\hypersetup{hidelinks}
\usepackage{url}
\usepackage{booktabs}
\usepackage{amsmath,amssymb}
\usepackage{graphicx}
\usepackage{microtype}

\title{L2R-EV: Learning What to Repair in Electric Ride-Pooling 
with Finite Charger Queues}
\author{%
  Mai Pham \\
  Thayer School of Engineering \\
  Dartmouth College \\
  \texttt{mai.p.pham.th@dartmouth.edu}
  \and
  Vikrant S. Vaze \\
  Thayer School of Engineering \\
  Dartmouth College \\
  \texttt{vikrant.s.vaze@dartmouth.edu}
  \and
  Peter Chin \\
  Thayer School of Engineering \\
  Dartmouth College \\
  \texttt{peter.chin@dartmouth.edu}
}
\begin{document}
\maketitle

\begin{abstract}
Electric ride-pooling must jointly manage passenger service, routes, batteries, and finite chargers; a locally useful relocation can reduce later service. We introduce a discrete-event ride-pooling simulator with ordered passenger stops, pickup and ride-time constraints, battery reserves, charger travel, and finite first-come, first-served (FCFS) charging queues. On top of a common dispatcher, we add a learned repair layer that moves uncollected passengers between vehicles, uses a supervised score to estimate later effects, and uses a two-stage proximal-policy-optimization (PPO) policy to focus computation on promising moves. Every executed move must satisfy route, battery, and time constraints and provide an immediate route-cost improvement. For larger instances, we limit the number of candidate moves and apply several non-conflicting repairs at once. On 12 held-out Manhattan episodes with 100 requests, 30 EVs, and one plug per station, supervised repair reduces operational cost by 6.57\% versus no repair (12/12 wins) and serves 74.92 rather than 72.75 requests. Two-stage PPO stays within 1.43\% of supervised repair with 95.4\% fewer exact trials, while our experiments scale to 8,000 requests and 2,400 EVs with cost reductions across all three larger development settings. Queue-aware charging reduces waiting by 41.0\%; under a declared 1.5-kW queue-idle load, operational electricity per served request falls 1.73\%.
\end{abstract}

\section{Introduction}

Ride-pooling can serve compatible trips together and thereby raise vehicle occupancy \citep{santi2014pooling}; electrification also removes tailpipe emissions. Neither change guarantees lower system-wide emissions. Empty travel, electricity supply, induced demand, and shifts from public transport remain consequential \citep{ipcc2022transport,erhardt2019congestion}. Operational evaluation must therefore connect passenger service to distance, energy, and charging rather than treating electrification alone as the outcome.

The operational problem is difficult for a second reason: it is dynamic. Requests arrive while vehicles already carry ordered pickup and drop-off commitments. Every update must preserve passenger time limits, capacity, battery reserve, travel to chargers, and competition for finite plugs. A move
that shortens two routes now can place scarce vehicle capacity far from future demand or expose a low-charge vehicle to congestion. We therefore estimate a move's effect on later service instead of accepting every immediate saving.

Prior work provides important pieces of this problem. Dynamic trip--vehicle assignment constructs feasible high-capacity pooled trips whenever new requests are processed \citep{alonso2017ondemand}. Dynamic electric ride-hailing learns estimates of future reward for dispatch, charging, and repositioning \citep{kullman2022dynamic}, while EV-routing research documents the additional
battery and charging decisions \citep{pelletier2016electric}. Learning what to defer shows how reinforcement learning can focus computation on selected parts of a very large problem \citep{ahn2020learning}. The open intersection is a dynamic, high-capacity pooled service with explicit shared-plug queues, where learned guidance must remain compatible with exact route feasibility.

Capacitated EV routing can coordinate planned routes so that they respect the number of plugs at each station \citep{froger2022capacitated}. Online pooling adds a different coupling: request assignment changes which EV will reach a charger, its arrival time, its required energy, and whether passenger dispatch interrupts charging. For this setting, the remaining gaps are to estimate the future service effect of a feasible route repair, restrict exact checks without discarding valuable
interactions, and preserve end-to-end execution as demand and fleet size grow.

We make three contributions. First, we provide a dynamic EV ride-pooling simulator that advances across request, route, and charging events and models finite charger queues. Second, we combine learned request reassignment with exact passenger and battery checks; reinforcement learning can further choose where to spend this budget. Third, we evaluate service, cost, charger congestion, computation, and operational electricity, including dynamic execution through 8,000 requests.

\section{Dynamic EV simulator}
\begin{figure}[t]
\centering
\includegraphics[width=\linewidth]{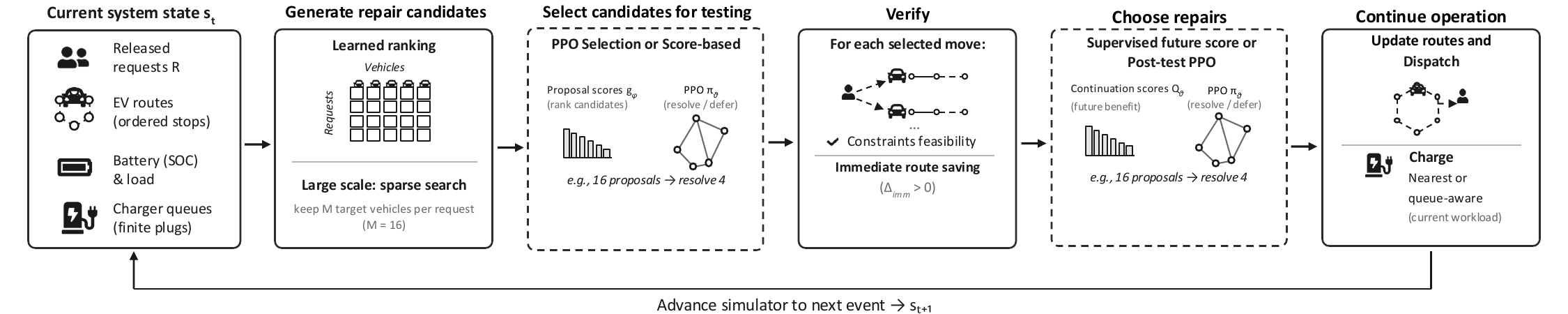}
\caption{L2V-EV Framework. Learning narrows the search; the simulator verifies executed
repair.}
\end{figure}

At time $t$, the observable state $s_t$ contains released requests, ordered stops, vehicle positions and loads, SOC, charger state, and current station queues. Requests and routes are subject to maximum pickup wait, maximum ride time, passenger-capacity, SOC, and reserve constraints. We
minimize the simulator's composite operational cost
\begin{equation}
 C=\alpha_{\rm rej}N_{\rm rej}+\alpha_{\rm veh}T_{\rm veh}
   +\alpha_{\rm en}E+\alpha_{\rm wait}T_{\rm wait}
   +\alpha_{\rm exc}T_{\rm excess},
 \label{eq:objective}
\end{equation}
where rejection, vehicle time, energy, passenger wait, and excess ride time have fixed weights. Lower is better. Charger queue time is a reported mechanism metric; it is not silently added to Eq.~\ref{eq:objective}.

\paragraph{Finite charger queues}

Each station has $m$ identical plugs. Arriving vehicles join a deterministic first-come, first-served queue. A waiting vehicle receives no energy. A vehicle releases its plug as soon as charging ends or passenger dispatch makes it leave. Charging completion is an event even when no passenger arrives at that time.
% The comparison methods include greedy insertion, which adds each new request to the cheapest feasible route, and rolling reoptimization, which repeatedly rebuilds the current assignment. Charging uses either the nearest station or the station with the smallest travel time plus visible workload. A perfect information reference is used on small cases; it sees future requests in advance but ignores competition for plugs. 
When two policies are compared, they receive the same request stream and initial fleet. Appendix~\ref{app:environment} gives the full
event rules, weights, data, and benchmarks.

\section{L2R-EV: Learning What to Repair in Electric Ride-Pooling}

A \emph{repair} is a reassignment of an accepted passenger before pickup. Our method first proposes promising reassignments cheaply, it then pays for an exact simulator test of selected proposals, then it accepts only tested moves that improve the current routes and are predicted to help later service.

\paragraph{Repair and exact evaluation.}
Formally, a repair $a=(v_{\rm src},r,v_{\rm dst})$ moves request $r$ from its current vehicle to another vehicle. The simulator tries the removal and reinsertion, recomputes both routes, and checks every constraint. The move proceeds only if it is feasible and reduces the remaining cost of these two routes. Rejected tests restore the original routes and event times exactly.

\paragraph{Learning the future effect of a repair.}
Before the exact test, a small neural network $g_\phi(s,a)$ ranks proposals by their likely immediate saving; at each repair time, at most $K=96$ proposals are tested. A second network $Q_\theta(s,a)$ estimates whether a tested move also improves the later episode. We train it from paired simulations that start from the same state and see the same future requests: one accepts the move and the other does nothing. Their later cost difference is the training label. This future-score network is supervised; PPO below is the reinforcement-learning component. The networks observe request and vehicle location, remaining time before route constraints are violated, load, SOC, reserve, known demand, and visible charger workload. They never observe unreleased requests. The method accepts the best feasible move only when its predicted future effect exceeds a threshold
chosen on validation data. Appendix~\ref{app:learning} gives the architecture, loss, label horizon, and feature list.

\paragraph{Learning which proposals to test.}
Exact testing remains costly. We therefore train a policy with proximal policy optimization (PPO) \citep{schulman2017ppo}. Here \emph{resolve} means paying for an exact test, while \emph{defer} means leaving that proposal untested at the current repair time. The policy sees 16 proposals and resolves four. After each test, a second decision may reject the move using its now-known route saving. Training uses final episode cost so the policy can value later passenger and charging effects. Appendix~\ref{app:deferral} gives the PPO objective, network, and reward assignment.

\paragraph{Sparse scale extension.}
Testing every request with every vehicle creates $R\!\times\!V$ proposals, where $R$ is the number of assigned but unpicked requests. At large scale, each request instead keeps $M=16$ vehicles: ten selected by learned request--vehicle similarity, four nearby vehicles, and two random vehicles for exploration. This heuristic makes candidate generation grow as $MR$. We can also execute several tested
moves together when no vehicle appears in more than one move. The maximum batch size is $\lceil V/150\rceil$, giving 4, 8, and 16 simultaneous repairs for fleets of 600, 1,200, and 2,400 EVs. Appendix~\ref{app:learning} describes the sparse search and batching rules.

\paragraph{Charging and operational energy.}
Low-SOC EVs choose either the nearest charger or the station with the smallest travel time plus current workload. Workload is the remaining charge time of vehicles already charging or waiting.
For the climate sensitivity, queue-idle auxiliary energy is $E_{\rm aux}=P_{\rm aux}T_{\rm queue}$, where $T_{\rm queue}$ is measured in hours and the  auxiliary load is $P_{\rm aux}=1.5$ kW.

\section{Results}

% The main setting draws Manhattan demand from aggregate tables containing 110.9 million trips. Each episode has 100 requests, 30 EVs, four passenger seats per EV, a 1.5-hour operating window, and one plug per station. We use 12 independent held-out random seeds and compare methods on the same seed.
% Appendix~\ref{app:protocol} gives the statistical tests and all controls.
The main experiment (Table~\ref{tab:repair4}) uses 12 paired, held-out Manhattan episodes, each with 100 requests, 30 four-seat EVs, a 1.5-hour window, and one plug per station; the source tables represent 110.9 million trips. In Table~\ref{tab:repair},
\emph{none} disables repair, \emph{rolling reopt.} rebuilds assignments, \emph{random immediate} samples up to $K$ repair candidates and commits the
feasible move with the largest exact route saving, \emph{quota random} samples four of the same 16 proposals, \emph{proposal score} tests the four highest $g_\phi$ scores, and \emph{before/after} adds post-test PPO. Appendix~\ref{app:protocol} gives exact implementations and statistical tests.

% \begin{table}[h]
% \centering
% \caption{Live EV repair on 12 held-out episodes. Cost reduction and wins are relative to no repair; tests saved are relative to supervised repair.}
% \label{tab:repair4}\label{tab:repair}
% \small\setlength{\tabcolsep}{4.2pt}
% \begin{tabular}{lrrrrrr}
% \toprule
% Method & cost $\downarrow$ & reduction & wins & served & tests/ep. $\downarrow$ & tests saved\\
% \midrule
% No repair          & 148,914 & ---    & ---   & 72.75 & 0     & ---\\
% PPO before test    & 144,178 & 3.18\% & 9/12  & 73.67 & 70    & 95.4\%\\
% PPO before/after   & 141,121 & 5.23\% & 8/12  & 74.58 & 70    & 95.4\%\\
% Supervised future score & 139,135 & 6.57\% & 12/12 & 74.92 & 1,530 & 0\%\\
% \bottomrule
% \end{tabular}
% \end{table}

\begin{table}[h]
\centering
\caption{Live EV repair comparisons. Each entry reports candidate/control;
effects and wins favor the candidate. The final comparisons use separate
12-seed blocks. Exact tests count relocation trials;
``--'' denotes a dispatcher without a repair layer.}
\label{tab:repair4}\label{tab:repair}
\scriptsize\setlength{\tabcolsep}{2.5pt}
\begin{tabular}{lrrrrr}
\toprule
Candidate / control & cost & effect & wins & served & exact tests\\
\midrule
\multicolumn{6}{l}{Shared held-out repair block}\\
PPO before test / none & 144,178/148,914 & 3.18\% & 9/12 & 73.67/72.75 & 70/0\\
PPO before/after / none & 141,121/148,914 & 5.23\% & 8/12 & 74.58/72.75 & 70/0\\
Supervised / none & 139,135/148,914 & 6.57\% & 12/12 & 74.92/72.75 & 1,530/0\\
Supervised / rolling reopt. & 139,135/185,653 & 25.06\% & 11/12 & 74.92/63.92 & 1,530/--\\
\midrule
\multicolumn{6}{l}{Repair-heuristic block}\\
Future score / random immediate & 136,265/138,324 & 1.49\% & 5/12 & 76.96/76.33 & 1,564/1,564\\
\midrule
\multicolumn{6}{l}{Matched-work allocation block}\\
PPO / quota random & 176,681/186,836 & 5.43\% & 9/12 & 91.33/89.42 & 71.0/71.7\\
PPO / proposal score & 176,681/183,924 & 3.94\% & 9/12 & 91.33/90.00 & 71.0/71.0\\
\bottomrule
\end{tabular}
\end{table}

% \paragraph{Solution quality and exact tests.}
% The supervised future score improves cost on all 12 seeds (95\% interval $[3.91,10.58]\%$, $p=0.00049$). In a separate 12-episode comparison using the same number of tests, PPO improves 5.43\% over choosing four proposals randomly and 3.94\% over choosing the four highest immediate-saving predictions. Notably, using PPO both before and after the test gives cost within 1.43\% of supervised repair while reducing exact evaluations from 1,530 to 70 per episode.

\paragraph{Solution quality and exact tests.}
The supervised future score improves cost on all 12 seeds (95\% interval $[3.91,10.58]\%$, $p=0.00049$). At matched work, PPO improves over both random and proposal-score allocation. Using PPO before and after testing gives cost within 1.43\% of supervised repair while reducing exact evaluations from 1,530 to 70 per episode.
Against rolling reoptimization with the same workload-aware charging rule,
supervised repair lowers aggregate cost by 25.06\%, wins 11/12 episodes, and
serves 11.0 more requests on average (95\% interval $[13.56,29.75]\%$).

\begin{table}[h]
\centering
\caption{The same supervised repair model across charger capacities. Entries are episode means; queue s is waiting under repair. Rows with 2--8 plugs use four episodes and the other rows use 12.}
\label{tab:capacity4}\label{tab:capacity}
\small\setlength{\tabcolsep}{2.5pt}
\begin{tabular}{lrrrrrrr}
\toprule
Plugs & episodes & no repair & repair & reduction & wins & extra served & queue sec.\\
\midrule
Unlimited & 12 & 151,560 & 141,027 & 6.95\% & 11/12 & +2.00 & 0\\
1         & 12 & 148,914 & 139,135 & 6.57\% & 12/12 & +2.17 & 2,746\\
2         & 4  & 167,785 & 156,689 & 6.61\% & 3/4   & +2.00 & 238\\
4         & 4  & 167,377 & 150,913 & 9.84\% & 3/4   & +3.25 & 0\\
8         & 4  & 167,377 & 150,913 & 9.84\% & 3/4   & +3.25 & 0\\
\bottomrule
\end{tabular}
\end{table}
\vspace{-7pt}

\paragraph{Finite-capacity robustness.} 
The same trained model reduces cost from one plug through unlimited charging. Four and eight plugs produce identical routes and zero waiting, confirming that these settings provide enough plugs. With one plug, repair still wins on all 12 paired episodes while operating through the FCFS queues (Table~\ref{tab:capacity4}).

\begin{table}[h]
\centering
\caption{Scaling with simultaneous nonconflicting repairs. Cost reduction and wins use paired no repair. Test fraction is the percentage of all $R\!\times\!V$ request--vehicle combinations tested exactly.}
\label{tab:scale4}\label{tab:dynamic-scale}
\scriptsize\setlength{\tabcolsep}{2.5pt}
\begin{tabular}{lrrrrrrrr}
\toprule
$(n,V)$ & runs & batch & reduction & wins & extra served & proposed & tested & test frac.\\
\midrule
$(100,30)$    & 12 & 1  & 6.57\% & 12/12 & +2.17 & 3,069   & 1,530 & ---\\
$(2000,600)$  & 4  & 4  & 2.82\% & 4/4   & +18.5 & 35,880  & 1,748 & 0.130\%\\
$(4000,1200)$ & 4  & 8  & 0.83\% & 3/4   & +9.0  & 72,048  & 1,744 & 0.032\%\\
$(8000,2400)$ & 2  & 16 & 1.69\% & 2/2   & +55.0 & 142,784 & 1,736 & 0.008\%\\
\bottomrule
\end{tabular}
\end{table}
\vspace{-7pt}

\paragraph{Large-scale experiments.}
Our large-scale configuration combines sparse $M=16$ candidate retrieval, the supervised future score, and vehicle-disjoint repair batches (Table~\ref{tab:scale4}). It lowers cost relative to the same dispatcher without repair in 9 of 10 episodes, and relative to the same sparse repair method restricted to one move per epoch in all 10. The share of request--vehicle combinations proposed falls from 2.67\% to 0.67\%, and the share tested exactly falls from 0.130\% to 0.008\%. Exact work stays near 1,700 tests while the number of requests grows fourfold from 2,000 to 8,000.

\paragraph{Queue waiting and operational energy.}
Current-workload charging cuts queue time by 41.0\% versus nearest charging with greedy dispatch and 36.0\% with reoptimization. At 1.5 kW, operational electricity per served request falls 1.73\% (0.642 versus 0.653 kWh; $[0.54,2.98]\%$, $p=0.0186$) \citep{doe2024coldbev}. Applying the Environmental Protection Agency's
New York City/Westchester average factor gives the same operational CO$_2$e percentage \citep{epa2025egrid}. Appendix~\ref{app:climate} reports traction-energy, terminal-SOC, grid-factor, and lifecycle scope checks.

\paragraph{Ablation Studies.}
Appendix D separately removes learned proposal ordering, future-score commitment, PPO allocation, and multi-move batching under matched budgets.

\section{Conclusion and outlook}

Exact simulation preserves feasibility, supervised repair improves held-out EV
service over no repair and rolling reoptimization, PPO reduces exact testing,
and sparse vehicle-disjoint repair completes dynamic episodes with 8,000 requests. Future work will freeze and confirm the large-scale rules across more datasets. It will also train batch-aware future scores and charger policies that value downstream passenger service. Richer energy physics and time-varying grid emissions are needed to connect these operational gains to climate outcomes.

\clearpage
\bibliographystyle{unsrt}
\bibliography{references}

\clearpage
\appendix
\section{Environment, data, and comparison methods}
\label{app:environment}

\paragraph{Events and state.}
The simulator advances directly to the next request release, passenger-stop
arrival, charging completion, or scheduled decision time. Its observable state
contains all released requests; each vehicle's position, ordered stops, load,
and state of charge (SOC); and the vehicles charging or waiting at every
station. Unreleased requests are not visible to a policy.

An action may assign or reject a new request, reassign an accepted passenger
before pickup, reposition an idle vehicle, or send a low-SOC vehicle to charge.
The simulator then advances all vehicles and chargers to the earliest next
event.

\paragraph{Passenger and battery constraints.}
A vehicle route is an ordered sequence of pickup and drop-off stops. Insertion
and repair enforce a 300-second pickup deadline, a maximum ride time equal to
twice that request's direct travel time, the stated passenger capacity, and a
5-kWh battery reserve. A passenger already inside a vehicle cannot be moved.
The EV model uses a 62-kWh battery, constant traction consumption of
0.15 kWh/km, and 72-kW charging up to the battery limit.

\paragraph{Operational objective.}
The cost in the main paper uses
$(\alpha_{\rm rej},\alpha_{\rm veh},\alpha_{\rm en},
\alpha_{\rm wait},\alpha_{\rm exc})=(5000,0.5,0.1,1,2)$.
Time coefficients are per second and energy is measured in kWh. A rejected
request therefore costs 5000; vehicle time, energy, passenger waiting, and
ride time beyond direct travel contribute the remaining terms. Charger-queue
waiting is reported separately and is not added to this objective. This
separation lets us test whether reducing congestion also improves passenger
service rather than assuming that it does.

\paragraph{Finite charging queues.}
Each station has $m$ identical plugs and a deterministic first-come,
first-served queue. Arrival time determines queue order, with vehicle ID used
only to break simultaneous-arrival ties. Waiting vehicles receive no energy.
A vehicle releases its plug when charging ends or passenger dispatch makes it
leave; the first waiting vehicle enters immediately. We record queue waiting,
active charging, queue entries, maximum queue length, plug use, charged energy,
traction energy, and vehicle distance.

The workload-aware charging rule minimizes travel time plus the unfinished
charging work already visible at a station. It includes vehicles charging,
vehicles waiting, and earlier charging choices in the same fleet action. It
does not predict future requests or future charger arrivals. With unlimited
plugs, it reduces exactly to nearest-station charging.

\paragraph{Demand and travel data.}
The Manhattan inputs are the aggregate 2018 weekday tables distributed with
the pyhailing benchmark \citep{kullman2022dynamic}. The demand table represents
110,932,862 trips in 1,600,497 weekday--15-minute--origin--destination rows.
It covers five weekdays, 96 time bins, 61 taxi zones, and 3,720 observed
directed zone pairs. A second table provides 357,216 zone-pair/time speed
records; its median speed is 20.9 km/h. The charging geometry contains 302
Manhattan parking-lot locations.

An episode samples a weekday and a contiguous 1.5-hour window. It draws
origin--destination rows according to their historical trip counts, samples
endpoints inside the selected zones, and computes street-grid-aligned Manhattan
distance with the empirical time-varying speeds. Vehicles begin at charging
locations using the benchmark's initialization weights. Unless full SOC is
specified, initial SOC is uniform on $[0,62]$ kWh. The simulator samples
individual requests from aggregate counts; it does not replay identified taxi
trips.

In the held-out $n=100$ experiment, an episode-level demand multiplier creates
realistic volume variation, so $n$ is a nominal mean and realized demand ranges
from 42 to 123. The large scaling experiments disable this multiplier and
generate exactly the stated $n$, isolating computational scale from demand
variation.

\paragraph{Comparison methods.}
Greedy insertion assigns each arriving request to its cheapest feasible route.
Rolling reoptimization periodically rebuilds the assignment from the requests
and vehicles currently available. Each dispatcher is paired with nearest or
workload-aware charging. On small cases, the perfect-information reference
sees future requests but does not schedule competition for plugs; it is
therefore a capacity-relaxed reference, not a queue-aware optimum. Policies in
a pair receive identical demand, initial vehicle positions, and initial SOC,
then evolve independently after their actions differ.

The main repair table uses the following labels. \emph{No repair} retains
greedy insertion and workload-aware charging but disables passenger
reassignment. \emph{Supervised repair} ranks candidates with $g_\phi$, applies
the exact feasibility and immediate-saving test, and commits a move only when
the supervised future score $Q_\theta$ also passes its validation threshold.
\emph{PPO before test} uses PPO to choose which proposals receive exact tests,
then commits every selected, conflict-free proposal that passes the exact
test. \emph{PPO before/after} adds a second PPO decision that may abstain after
the exact saving is known. Rolling reoptimization uses the same workload-aware
charging rule as the repair arms but has no relocation layer.

\section{Learned repair: architecture and training}
\label{app:learning}

\paragraph{Candidate and exact test.}
A repair candidate $a=(v_s,r,v_t)$ moves an assigned but unpicked request $r$
from source vehicle $v_s$ to target vehicle $v_t$. For every selected
candidate, the evaluator snapshots the two stop lists, request ownership,
schedules, and next-event times. It removes the request, finds its best
feasible insertion into the target route, and computes the resulting change in
the remaining cost of both routes. A candidate can be executed only when it
passes every passenger and battery constraint and has positive immediate route
saving. If it is rejected, the complete pre-test state is restored. A
regression test verifies that evaluating candidates without executing one
leaves the next event unchanged.

\paragraph{Proposal and future-score networks.}
The proposal network $g_\phi(s,a)$ ranks candidates before exact testing. The
future-score network $Q_\theta(s,a)$ decides whether a tested, immediately
improving move is likely to help the rest of the episode. The notation
$Q_\theta$ denotes a supervised continuation score, not a Q-learning update.

Separate two-layer multilayer perceptrons encode request and vehicle features
into 64-dimensional vectors. For request representation $h_r$, source and
target representations $h_s,h_t$, and mean request and vehicle context
$\bar h_r,\bar h_v$, the interaction head receives
$[h_r,h_s,h_t,h_t-h_s,\bar h_r,\bar h_v]$. Features describe geometry, pickup
and ride-time slack, route completion time and location, load, SOC, reserve,
released-demand density, charging state, selected-station delay, and visible
network workload.

\paragraph{Paired-rollout labels.}
Training labels use two copies of the same state and the same already-sampled
future requests. One copy executes the candidate and the other does not; both
then use greedy dispatch and workload-aware charging for at most 24 simulator
steps or 1,800 seconds, whichever comes first. Denote this horizon by $H$.
For rollout utility $U_H$ and state-dependent scale $S(s)$, the label is
$y(s,a)=[U_H(s\!\xrightarrow{a})-U_H(s\!\xrightarrow{\varnothing})]/S(s)$.
Only feasible candidates with positive immediate saving receive this label.
Training combines a Smooth-L1 regression loss with a pairwise ranking loss:
\[
 \mathcal L=\operatorname{SmoothL1}(Q_\theta,100y)
 +|\mathcal P|^{-1}\!\sum_{(i,j)\in\mathcal P}
 w_{ij}\log\!\left(1+e^{-(Q_i-Q_j)}\right),
\]
where $\mathcal P=\{(i,j):y_i>y_j\}$ and $w_{ij}$ is a clipped normalized
label difference. Data aggregation includes states reached after previous
repairs. The commit threshold is selected on validation data, and abstention
is allowed.

\paragraph{Sparse candidate retrieval.}
Complete candidate generation requires $R V$ request--target pairs when $R$
accepted passengers remain unpicked. At large scale, a dual encoder maps each
request--source pair to a 32-dimensional query and each vehicle to a key. It is
distilled on small snapshots where complete interaction scores are available.
A $k$-d tree returns ten learned targets per request; four nearby and two
uniformly sampled targets add geometric and exploratory coverage. Thus each
request retains $M=16$ targets without constructing the full interaction
matrix. Retrieval only proposes candidates: every selected move still passes
the same exact insertion and future-score checks.

\paragraph{Simultaneous repairs.}
At the $n=100,V=30$ scale, at most one repair is executed at each repair epoch.
For larger fleets, accepted candidates are sorted by score and a compatible
batch is executed. Compatibility requires that no vehicle appears in more than
one move, so the checked route changes do not overlap. The scale experiments
use the fixed rule $\lceil V/150\rceil$, which gives batch limits 4, 8, and 16
for 600, 1,200, and 2,400 vehicles. All settings retain the same per-epoch
exact-test cap $K=96$.

\subsection{Architecture and training hyperparameters}
\label{app:hyperparameters}

Table~\ref{tab:network-hyperparameters} records the deployed neural
architectures. Here ``features'' gives the request and vehicle input widths;
$d$ is the hidden width. All multilayer perceptrons use GELU activations and
LayerNorm after their first hidden layer. The proposal and future-score models
share the interaction form described above but have separate weights and
feature contracts. The PPO policy does not share weights with either
supervised scorer.

\begin{table}[h]
\centering
\caption{Deployed architecture hyperparameters. ``MLP widths'' excludes the
input layer and lists the learned output dimensions.}
\label{tab:network-hyperparameters}
\small
\setlength{\tabcolsep}{4pt}
\begin{tabular}{p{0.19\linewidth}p{0.18\linewidth}p{0.38\linewidth}p{0.13\linewidth}}
\toprule
Component & Input features & MLP widths & Output/use \\
\midrule
Proposal scorer & 12 request, 16 vehicle & encoders: $64,64$; interaction: $128,1$ & candidate rank \\
Future scorer & 16 request, 28 vehicle & encoders: $64,64$; interaction: $128,1$ & commit score \\
Sparse retriever & 16 request, 28 vehicle & encoders: $64,64$; query: $128,32$; key: $64,32$ & normalized 32-D keys \\
Pre-exact PPO & 17 per candidate & item: $64,64$; pooled head: $64,2$; value: $64,1$ & resolve/defer \\
Post-exact PPO & 17 plus exact gain & head: $64,2$ & commit choice \\
\bottomrule
\end{tabular}
\end{table}

Table~\ref{tab:training-hyperparameters} gives the corresponding training and
deployment settings. Supervised networks use AdamW and PPO uses Adam; gradients
are clipped to norm 1.0. The supervised checkpoints use early stopping with
patience 25, so 120 is a maximum rather than necessarily the selected epoch.

\begin{table}[h]
\centering
\caption{Training and deployment hyperparameters for reported checkpoints.}
\label{tab:training-hyperparameters}
\small
\setlength{\tabcolsep}{4pt}
\begin{tabular}{p{0.19\linewidth}p{0.12\linewidth}p{0.13\linewidth}p{0.46\linewidth}}
\toprule
Component & Learning rate & Batch & Other fixed settings \\
\midrule
Proposal scorer & $3\!\times\!10^{-4}$ & 8 states & 120 epochs; weight decay $10^{-4}$; ranking weight 1 \\
Future scorer & $3\!\times\!10^{-4}$ & 4 states & 120 epochs; weight decay $10^{-4}$; 24-step/1,800-s paired rollout; regression/ranking weights 1/1; validation threshold 1.0 \\
Sparse retriever & $3\!\times\!10^{-4}$ & 4 states & 120 epochs; weight decay $10^{-4}$; temperature 1; label weight 0.25; $10+4+2$ learned/nearest/random targets \\
Pre-exact PPO & $3\!\times\!10^{-4}$ & 4 episodes & 24 episodes; four PPO passes; clip 0.2; entropy 0.03; value weight 0.5; 16 proposals, 4 tests \\
Post-exact PPO & $10^{-3}$ & 4 episodes & 8 additional episodes; four PPO passes; clip 0.2; entropy 0.03; value weight 0.5 \\
\bottomrule
\end{tabular}
\end{table}

\section{PPO allocation of exact tests}
\label{app:deferral}

\paragraph{Action and observations.}
The pre-exact policy receives 16 cheap proposals and selects four for exact
testing without replacement. Its 17 causal features describe the proposal
score, source--target and target--pickup distances, time slack, trip duration,
source and target SOC and load, episode time, and fleet and queue summaries. A
shared item network combines each proposal with the mean candidate-set context
and outputs resolve/defer logits. Conflicting choices are removed before a
batch is executed. Every resolved proposal must still pass the exact
feasibility and immediate-saving test.

\paragraph{Training.}
The selector is trained for 24 live episodes with clipped PPO
\citep{schulman2017ppo}. Its policy loss is
\[
 -\mathbb E_t\!\left[\min\!\left(r_t\hat A_t,
 \operatorname{clip}(r_t,1-\epsilon,1+\epsilon)\hat A_t\right)\right],
\]
plus value and entropy terms. Each rollout is paired with a same-seed policy
that retrieves the same proposals but selects its four exact tests uniformly.
This random policy is a variance-reduction baseline, not a teacher. Advantages
are aligned by repair epoch and selection round and normalized over a
four-episode rollout batch. Final episode cost supplies the learning signal.

A second PPO head can act after exact evaluation. It observes the same causal
features plus the known immediate route saving and chooses commit or abstain.
It is trained for eight additional episodes while the pre-exact selector is
frozen. Neither PPO head uses the supervised continuation label.

\paragraph{Matched-work evaluation.}
Table~\ref{tab:ppo-appendix} compares the frozen pre-exact policy with two
controls on 12 episodes. All methods retrieve 16 proposals and perform four
exact tests, so differences cannot be explained by unequal solver work. The
first six episodes were inspected before a contiguous six-episode extension;
the pooled $p$-values are therefore descriptive rather than confirmatory.

The \emph{quota-random} control samples four distinct proposals uniformly
without replacement from the same set of 16; two random-selector streams are
averaged within each episode. The \emph{proposal-score} control sorts that
same set by the cheap proposal score $g_\phi(s,a)$ and tests the top four.
Both controls use the same conflict cleanup, exact evaluator, positive-saving
requirement, dispatcher, and charging rule as PPO. Thus this comparison changes
only how the four-test budget is allocated.

\begin{table}[h]
\centering
\caption{Pre-exact PPO relative to matched-work controls at $n=100,V=30$.
Positive effect means lower cost with PPO.}
\label{tab:ppo-appendix}
\small
\begin{tabular}{lrrrr}
\toprule
Control & effect & wins & bootstrap 95\% & sign-flip $p$\\
\midrule
Quota random & 5.43\% & 9/12 & $[1.48,10.72]\%$ & 0.0317\\
Proposal score & 3.94\% & 9/12 & $[1.85,6.34]\%$ & 0.0083\\
\bottomrule
\end{tabular}
\end{table}

On the 12 held-out seeds used in the main repair table, pre-exact PPO performs
69.7 exact tests per episode and reduces cost by 3.18\% relative to no repair.
Adding the post-exact head raises the reduction to 5.23\% at the same exact
work. Supervised repair reduces cost by 6.57\% but performs 1,530.3 tests. The
two-stage PPO point is therefore 1.43\% more costly in aggregate while using
95.45\% fewer exact tests.

\section{Component ablations}
\label{app:ablations}

We evaluate components by changing one decision rule while keeping the
dispatcher, charger, repair epochs, candidate set, and exact-test cap fixed.
Table~\ref{tab:repair-factorial} crosses two choices. The proposal order is
either learned or uniformly random. After exact feasibility and immediate
route saving are known, commitment either accepts the best immediate saving or
uses the supervised future score with abstention.

\begin{table}[h]
\centering
\caption{Proposal-ordering $\times$ commitment ablation on 12 fresh paired
episodes. Effects and wins are relative to no repair.}
\label{tab:repair-factorial}
\small
\begin{tabular}{llrrrr}
\toprule
Proposal order & commitment & cost $\downarrow$ & effect & wins & exact tests\\
\midrule
None & none & 142,496 & --- & --- & 0\\
Random & immediate saving & 138,324 & 2.93\% & 7/12 & 1,564\\
Learned & immediate saving & 136,361 & 4.31\% & 7/12 & 1,550\\
Random & future score & 135,929 & 4.61\% & 9/12 & 1,561\\
Learned & future score & 136,265 & 4.37\% & 8/12 & 1,564\\
\bottomrule
\end{tabular}
\end{table}

Random ordering with future-score commitment has a positive paired interval
versus no repair ($[0.98,8.93]$\%, sign-flip $p=0.043$). The direct component
contrasts are less certain: learned versus random ordering changes aggregate
cost by +1.42\% with immediate commitment and $-0.25$\% with future-score
commitment; future-score versus immediate commitment changes cost by +1.73\%
under random ordering and +0.07\% under learned ordering. All four direct
component intervals cross zero. Thus the experiments support the complete
repair layer and show that future-aware abstention can work even with random
proposals, but they do not attribute the gain independently to learned proposal
ordering.

A second control uses the 12 principal held-out seeds and directly compares
learned future-score repair with a classical stochastic repair heuristic. The
heuristic samples up to $K=96$ candidates uniformly, evaluates them exactly,
and commits the feasible candidate with the largest immediate route saving.
It has mean cost 138,422 and performs 1,545 exact tests per episode; learned
future-score repair has mean cost 139,505 and performs 1,532 tests. The learned
method wins 8/12 episodes, but its mean episode-normalized effect is 0.01\%
with interval $[-5.32,4.57]\%$ and sign-flip $p=0.998$; its aggregate cost is
0.78\% higher. The immediate heuristic itself lowers aggregate cost by 7.05\%
versus no repair (10/12 wins). This matched-seed control therefore does not
support learned superiority over high-budget immediate-gain repair.

Two further matched-budget ablations isolate computation. First,
Table~\ref{tab:ppo-appendix} replaces PPO exact-test allocation with either a
quota-random allocator or deterministic proposal-score allocation; PPO lowers
cost by 5.43\% and 3.94\%, respectively. Second,
Table~\ref{tab:batch-ablation} replaces a vehicle-disjoint repair batch with a
single repair per epoch; batching wins all ten evaluated large-scale pairs.
The sparse retriever is required to avoid constructing the large all-pairs
matrix, but we do not have a feasible all-pairs control at $n=2000$--8000 and
therefore do not claim an isolated live-EV quality effect for retrieval.

The exact feasibility test is a safety requirement rather than an ablated
component: removing it would allow violations of passenger, capacity, and
battery constraints and would change the problem being solved.

\paragraph{Operational dispatcher benchmark.}
On the 12 principal held-out seeds, rolling reoptimization with the same
workload-aware one-plug charging rule has mean cost 185,653 and serves 63.92
requests. Supervised repair has mean cost 139,135 and serves 74.92. The paired
aggregate cost reduction is 25.06\%, with 11/12 wins, an episode-normalized
95\% interval of $[13.56,29.75]\%$, and exact sign-flip $p=0.00195$. The
reoptimization run independently reproduces the no-repair greedy means of
148,914 cost and 72.75 served, confirming that the episode configuration is
matched. Reoptimization is a dispatcher benchmark, not a component ablation.

\paragraph{Tractable perfect-information reference.}
We also solve four small cases with $n=20,V=8$ using a 60-second monolithic
mixed-integer programming limit. The mean certified lower bound is 4,846 and
the mean feasible perfect-information incumbent is 11,851; none of the four
brackets closes to proven optimality. The corresponding online greedy and
rolling-reoptimization costs are 39,737 and 62,130. Only the certified lower
bound is a rigorous bound for the finite-queue problem. The incumbent belongs
to a capacity-relaxed model that sees future requests but does not schedule
shared plugs, so it is reported as a small-case reference rather than a
queue-aware oracle.

\section{Experimental protocol}
\label{app:protocol}

\paragraph{Held-out EV experiment.}
The principal evaluation uses Manhattan demand, nominal $n=100$, $V=30$ EVs,
four seats, a 1.5-hour window, and one plug per station. The repair epoch is
300 seconds and the exact cap is $K=96$. Twelve seeds begin at 7,500,000 and
are disjoint from training and validation. The charger-capacity comparison
repeats no repair and supervised repair on these seeds with unlimited station
use. A smaller, shared four-seed extension tests two, four, and eight plugs.

The queue comparison uses 12 paired seeds beginning at 6,900,000. It compares
nearest and workload-aware charging with greedy insertion and rolling
reoptimization under unlimited and one-plug capacity.

\paragraph{Scaling protocol.}
The $n=2000,4000,8000$ experiments use fixed request counts, one plug per
station, the frozen $M=16$ retriever, and the same $K=96$ exact cap. For
bounded dispatch cost, each new request scans at most 64 vehicles and stops
after finding 16 feasible insertions. Mean travel times use a precomputed
$61^2\!\times96$ lookup instead of repeatedly joining the same speed table.
Unit and episode tests verify exact equality with the original lookup path.
The vehicle-disjoint repair-batch rule was selected after diagnosing the
failure of a fixed one-move limit, so these scale blocks are development
evidence.

\paragraph{Statistical unit and inference.}
The statistical unit is an independently seeded episode. Stochastic selector
streams are averaged inside an episode and are not counted as additional
samples. For candidate method $A$ and control $B$, episode improvement is
$100(C_i^B-C_i^A)/C_i^B$, so positive values favor $A$. We report the mean
paired effect, a 20,000-resample percentile bootstrap interval, paired wins,
and an exact two-sided sign-flip test. Policies in each pair share exogenous
episode draws.

\section{Additional operational evidence}

\paragraph{Robustness to charger capacity.}
The supervised repair model is frozen across all rows of
Table~\ref{tab:repair-capacity-energy}. It improves cost by 6.95\% with
unlimited plugs and 6.57\% with one plug on the 12 held-out seeds. The
two-, four-, and eight-plug rows use four shared seeds. Four and eight plugs
produce identical outcomes and no queueing, providing an abundant-capacity
null for that block.

\begin{table}[h]
\centering
\caption{Repair across charger capacities. Positive energy effect means lower
traction kWh per served request with repair.}
\label{tab:repair-capacity-energy}
\small
\begin{tabular}{lrrrrr}
\toprule
Plugs & episodes & cost effect & wins & served $\Delta$ & energy effect\\
\midrule
Unlimited & 12 & 6.95\% & 11/12 & +2.00 & +0.77\%\\
1         & 12 & 6.57\% & 12/12 & +2.17 & $-0.40$\%\\
2         & 4  & 6.61\% & 3/4   & +2.00 & $-0.30$\%\\
4         & 4  & 9.84\% & 3/4   & +3.25 & +2.03\%\\
8         & 4  & 9.84\% & 3/4   & +3.25 & +2.03\%\\
\bottomrule
\end{tabular}
\end{table}

\paragraph{Dynamic scaling.}
The main paper reports the scale-aware batches. Table~\ref{tab:batch-ablation}
shows the diagnosed bottleneck: when every epoch executes at most one repair,
the number of accepted moves remains about 18 even as the fleet grows. The
compatible batch uses the same scorer, sparse proposals, and exact-test cap,
but makes more of those already-checked moves useful. It improves over batch 1
on all ten evaluated large-scale episodes.

\begin{table}[h]
\centering
\caption{Scale-aware batch ablation. Effects are aggregate cost reductions.}
\label{tab:batch-ablation}
\small
\begin{tabular}{rrrrrr}
\toprule
$n$ & episodes & batch & batch 1 vs none & batch vs none & batch vs batch 1\\
\midrule
2,000 & 4 & 4  & +0.50\% & +2.82\% & +2.34\%\\
4,000 & 4 & 8  & $-0.99$\% & +0.83\% & +1.80\%\\
8,000 & 2 & 16 & +0.09\% & +1.69\% & +1.61\%\\
\bottomrule
\end{tabular}
\end{table}

Across $n=2000,4000,8000$, the exact-test fractions are 0.130\%, 0.032\%,
and 0.008\% of all request--vehicle interactions. Peak memory for learned
repair is 0.80, 1.24, and 2.33 GB. These small, post-hoc blocks establish
end-to-end dynamic execution and identify a useful batching mechanism; they do
not provide population-level estimates at these scales.

\paragraph{Static decomposition study.}
We separately test a three-layer, width-128 edge-conditioned GraphSAGE model
that assigns requests and vehicles to 12 balanced regions. Recursive exact
insertion begins in small leaves and promotes unresolved requests through
wider parent regions. Relative to flat global search, it lowers cost by
2.94\%, 2.05\%, and 0.96\% at $n=1000,2000,4000$ while using 39.2\%, 55.2\%,
and 67.6\% fewer exact insertions; it wins all 8, 4, and 3 paired instances.
At $n=8000$, one development instance reaches a 0.08\% cost gap with 68.7\%
fewer exact insertions and 3.16 times less solver time after widening the final
search cap. This static study tests the search structure, not dynamic EV
control.

\section{Distributional charging study}
\label{app:qplex}

QPLEX propagates discrete transient occupancy distributions from arrival and
service-time probability mass functions \citep{dieker2024qplex}. We use its
\texttt{StandardMultiserver} engine to compare shortlisted charging stations.
The arrival input combines station/time distributions fitted on eight
development episodes with Bernoulli arrivals from currently committed vehicle
routes. Charging-time distributions follow observable SOC deficits. For each
station, the controller compares its occupancy trajectory with and without the
candidate EV and adds the marginal expected buffer time to travel cost. The
event-driven simulator still executes the chosen station and its exact FCFS
queue.

This is a one-step distributional charging rule, not the full multi-step QDP
algorithm \citep{dieker2026qdp}. After a four-episode configuration gate, the
rule was frozen with five-minute periods, six-period lookahead, an eight-station
geometric shortlist, unit expected-buffer weight, and no tail penalty.

\begin{table}[h]
\centering
\caption{QPLEX charging study on 12 subsequent development episodes.}
\label{tab:qplex-gate}
\small
\begin{tabular}{lrrrr}
\toprule
Rule & cost $\downarrow$ & served & queue seconds $\downarrow$ & max queue\\
\midrule
Nearest & 155,082 & 77.33 & 4,307 & 1.58\\
Current workload & 155,102 & 77.33 & 2,493 & 1.25\\
QPLEX buffer & 152,338 & 78.08 & 2,104 & 1.17\\
\bottomrule
\end{tabular}
\end{table}

QPLEX lowers aggregate cost by 1.77\% versus nearest charging and 1.78\%
versus current workload, with 7/12 wins in each comparison. Queue waiting falls
51.1\% and 15.6\%, respectively. The bootstrap intervals cross zero. Eight
fresh seeds reproduce the favorable cost direction versus current workload
(3.53\%, 5/8 wins), but not a lower mean queue time. The engine itself uses
about 0.08 seconds per episode; total Python overhead is about 3.5\% relative
to the workload rule. These results support computational feasibility and
further frozen evaluation, not a superiority claim.

Route repair and QPLEX are not combined in the reported experiments. Repair
changes the charger-arrival process, so a reliable combination requires
policy-conditioned arrival distributions and a charging objective that values
future vehicle availability as well as queue occupancy.

\section{Operational energy and emissions boundary}
\label{app:climate}

\paragraph{Measured quantities.}
We record traction and charged energy, total, empty, and passenger-carrying
vehicle distance, passenger distance, occupancy, queue waiting, and terminal
SOC. Service-normalized quantities are computed within each episode before
averaging so that high-demand episodes do not receive extra statistical
weight.

On the 12 held-out repair episodes, no repair uses 45.50 kWh and 303.32 km to
serve 72.75 requests on average. Supervised repair uses 47.08 kWh and
313.89 km to serve 74.92. Energy per served request is 0.6247 versus 0.6272
kWh, an episode-normalized effect of $-0.35$\% with interval
$[-1.58,0.84]$\%. Thus the service gain is not also a demonstrated traction-
energy gain.

\paragraph{Queue-idle auxiliary load.}
Cabin conditioning can consume energy while an EV waits without moving. For
declared load $P_{\rm aux}$ and measured queue time $T_{\rm queue}$,
$E_{\rm aux}=P_{\rm aux}T_{\rm queue}$. Table~\ref{tab:aux-load} shows the
sensitivity of workload-aware versus nearest charging with greedy dispatch.
The 1.5-kW setting is close to cold-weather BEV cabin loads reported by the
U.S. Department of Energy \citep{doe2024coldbev}.

\begin{table}[h]
\centering
\caption{Operational kWh per served request under queue-idle auxiliary load
(12 paired episodes).}
\label{tab:aux-load}
\small
\begin{tabular}{rrrr}
\toprule
Auxiliary kW & nearest & workload-aware & reduction\\
\midrule
0.0 & 0.6250 & 0.6249 & 0.01\%\\
0.5 & 0.6343 & 0.6305 & 0.60\%\\
1.5 & 0.6531 & 0.6417 & 1.73\%\\
3.0 & 0.6812 & 0.6586 & 3.32\%\\
5.0 & 0.7186 & 0.6810 & 5.24\%\\
\bottomrule
\end{tabular}
\end{table}

At 1.5 kW, the paired bootstrap interval is $[0.54,2.98]$\% and the sign-flip
$p=0.0186$. The United States Environmental Protection Agency's Emissions \&
Generation Resource Integrated Database reports 865.7 lb CO$_2$e/MWh total
output for New York City and Westchester County, with 4.2\% gross grid loss
\citep{epa2025egrid}. Multiplying both policies by the corresponding
loss-adjusted factor preserves the 1.73\% operational CO$_2$e difference.

\paragraph{Accounting boundary.}
The simulator records energy charged at station $j$ and time bin $t$ as
$e_{jt}$, allowing operational accounting under a declared grid factor
$g_{jt}$ through $G_{\rm op}=\sum_{j,t}e_{jt}g_{jt}$. We retain an April-2025
New York Independent System Operator marginal-emissions profile for the New
York City zone for sensitivity analysis
\citep{nyiso2026imer}. However, finite episodes can end with different stored
battery energy. In the fresh QPLEX audit, the apparent 3.08\% reduction in
within-horizon charging CO$_2$ per served request coincides with 2.75\% lower
terminal battery energy. A terminal-neutral traction-plus-queue estimate is
only 1.15\% lower and has an interval crossing zero.

These operational calculations exclude vehicle and battery production,
charger construction, maintenance, vehicle lifetime, induced travel, modal
shift, and the passengers' counterfactual modes. We therefore report
operational mechanisms and sensitivity, not lifecycle or avoided-emissions
effects.

\section{Claim boundaries and reproducibility}
\label{app:claims}
\label{app:artifacts}

\paragraph{Supported conclusions.}
The held-out experiments support three principal conclusions: finite one-plug
queues bind; supervised repair improves the live EV objective and service; and
PPO provides a lower-exact-work repair point. The dynamic scale experiments
show complete execution through 8,000 requests and diagnose why
vehicle-disjoint repair batches are needed. The QPLEX and largest-scale results are
development evidence and are labeled as such.

The current evidence does not establish a dynamic global optimum, lifecycle or
causal avoided emissions, QPLEX/QDP superiority, learned decomposition
superiority over a matched random control, or population-level dynamic repair
effects at $n\geq2000$.

\paragraph{Canonical artifacts.}
The machine-readable index
\path{results/tccml2026_main_results.json} records every displayed value and
the raw file from which it was computed. The added same-seed dispatcher and
classical-repair controls are stored in
\path{results/ev_queue_reopt_sharedq_seeds12_s7500000.json} and
\path{results/ev_repair_classical_sharedq_seeds12_s7500000.json}. The
licensed small-case perfect-information run is stored in
\path{results/ev_pi_relaxed_small4_s7500000_licensed.json}. The repository
manifest further marks
artifacts as held-out, development, negative, or superseded. Environment and
queue logic are under \path{environment/}; learned policies are under
\path{agents/}; experiment and summary drivers are under \path{experiments/};
and immutable outputs are under \path{results/}. No result generated before
the exact transaction-restoration fix is used as paper evidence.

\end{document}